\documentclass[prl,twocolumn,aps,epsfig,graphicx,amssymb]{revtex4}
\usepackage{graphicx}
\begin{document}

\title{Stationary Pulses of Light in an Atomic Medium}
\author{M.~Bajcsy$^{1,2} $, A.~S.~Zibrov$^{1,3,4}$ and M.~D.~Lukin$^{1}$
}
\address{
$^1$Physics Department, Harvard University, Cambridge, MA 02138, USA
$^2$Division of Engineering and Applied Sciences, Harvard University, Cambridge, MA 02138, USA
$^3$Harvard-Smithsonian Center for Astrophysics, Cambridge, MA 02138, USA
$^4$Lebedev Institute of Physics, Moscow, Russia}



\maketitle

{\bf Physical processes that could facilitate  coherent control of light propagation are now actively explored
\cite{scullybook,harris-phystoday,lukin-nat,marlanrev,mabuchi}. In addition to fundamental interest, 
these efforts are stimulated by possibilities to develop, for example, a quantum memory for photonic states
 \cite{cirac97,kuzmich03,wal03}. At the same time, controlled localization and storage of photonic pulses 
may allow  novel approaches to manipulate light via enhanced nonlinear optical processes \cite{boyd}.  
Recently, Electromagnetically Induced Transparency (EIT) \cite{boller}  was used to reduce the group 
velocity of propagating light pulses \cite{hau99,kash99} and to reversibly map propagating light pulses 
into stationary spin excitations in atomic media \cite{fleischhauer00,liu01,phillips01,sashaback}.
 Here we describe and experimentally demonstrate a novel technique in which light propagating in  
a medium of Rb atoms is converted  into an excitation with localized, stationary electromagnetic energy, 
which can be held and released  after a controllable interval.  Our method creates  pulses of light with 
stationary envelopes bound to an atomic spin coherence, raising new 
possibilities for photon state manipulation and non-linear optical processes at low light levels.
 }

Several techniques to generate stationary light pulses in an atomic medium have  recently been proposed 
theoretically \cite{olga,axel}.  The present work is related conceptually to the ideas of Ref. \cite{axel},
but here we use a different approach involving spatial modulation of the absorptive properties of the medium.
The use of such a dissipative effect for coherent control of light is a novel phenomenon, closely related to 
earlier work on matched pulse propagation in EIT \cite{harris-phystoday,match}.
Although there exists a substantial body of  work on localization of light in random dielectrics 
\cite{gao}, Bragg gratings and photonic bandgap materials \cite{kogelnik,yablonovich,slusher}, the unique 
feature of the present method is that it allows one to accurately control the creation of a stationary light 
pulse and its release.

The present method can be understood  qualitatively by considering a
three-state "lambda" configuration of atomic states (Figure~1A). A large
ensemble of $N$ atoms is initially prepared in the  ground state $|g \rangle$. We use  forward (FD) 
and backward (BD) control beams, with time varying Rabi-frequencies
$\Omega_+(t)$ and $\Omega_-(t)$, respectively, to manipulate  a weak pulse of signal light. In our 
experimental realization the two control fields have  identical frequencies but opposite propagation  
directions.   The usual
EIT \cite{boller} corresponds to simultaneous propagation of the forward control and signal beams.
 When their frequency difference matches the level splitting between the ground state and a metastable 
("spin-flipped") state ($|s\rangle$) the medium becomes
transparent for the signal light, while the sharp atomic dispersion allows one to slow and localize an 
input signal pulse in the medium \cite{hau99,kash99}. By turning the control beam off while the pulse is 
in the medium \cite{fleischhauer00}, the signal amplitude vanishes while its state is stored in a 
stationary spin coherence. This atomic excitation can be converted back into  a light pulse, propagating 
in forward or backward direction, by application of the corresponding control beam 
\cite{liu01,phillips01,sashaback}.
The atomic coherence can be converted  into a stationary photonic excitation if the medium is illuminated 
simultaneously by FD and BD beams. Specifically, if the two create a standing wave 
pattern the EIT suppresses the signal absorption  everywhere but in the nodes of the  standing wave, 
resulting in a sharply peaked, periodic modulation of the atomic absorption for the signal 
light (Figure 1B). Illumination by these beams  also results in partial conversion of the stored atomic 
spin excitation into sinusoidally modulated signal light, but the latter cannot propagate in the medium due 
to Bragg reflections off the sharp absorption peaks, resulting in vanishing group velocity of the signal  pulse. 
Only after one of the control beams is turned off does the pulse acquire a finite  velocity and thus can  
leave the medium in the direction of the remaining  control beam.

To quantify these effects theoretically, consider interaction of atoms with resonant optical fields, 
represented by  plane waves. 
We decompose the signal field into  components propagating  in the forward
 and backward directions along the $z$ axis with wave-vectors $\pm k$ and  slowly varying amplitudes ${\mathcal E}_\pm$.
 Following   \cite{fleischhauer00,axel} we introduce  two components
$\Psi_\pm$ of a coupled excitation of light and an atomic spin wave ("dark-state polariton")
corresponding to forward and backward signal fields respectively.  In the experimentally relevant case of 
small group 
velocities the polariton components are represented by $\Psi_{\pm} = g\sqrt{N} {\mathcal E}_{\pm}/\Omega_{\pm}$, where $g$ is
the atom-field coupling constant \cite{scullybook}. Assuming further slowly varying  pulses  and negligible
 spin decoherence, we find that the components evolve according to
\begin{eqnarray}
{\partial \over \partial z}  \Psi_+ = -\alpha_- \xi (\Psi_+ - \Psi_-) - {1\over c} {\partial \over \partial \tau} (\alpha_+ \Psi_+ + \alpha_- \Psi_-), \\
{\partial \over \partial z}  \Psi_- = -\alpha_+ \xi (\Psi_+ - \Psi_-) + {1\over c} {\partial \over \partial \tau} (\alpha_+ \Psi_+ + \alpha_- \Psi_-)
\end{eqnarray}
and the spin coherence $S = N^{-1/2} (\alpha_+\Psi_+ + \alpha_-\Psi_-)$ with  $\alpha_\pm = |\Omega_\pm|^2/(|\Omega_+|^2 + |\Omega_-|^2)$. These equations describe two slow waves that are coupled
due to periodic modulation of atomic absorption  and group velocity.
The first term in the right hand side of Equations (1,2) is proportional to an absorption coefficient $\xi$ near resonant line center. When $\xi$ is large this term
gives rise to the pulse matching phenomenon  \cite{harris-phystoday,match}: whenever one of the fields is created the other will adjust itself within a short propagation distance to match its 
amplitude such that $\Psi_+-\Psi_-\rightarrow 0$. The scaled time, $\tau(t) = \int_{0}^t d t (|\Omega_+|^2 +|\Omega_-|^2)/g^2 N$, reflects the group velocity reduction associated with atomic dispersion.  
Finally, the center frequency of the signal light was chosen to
match its wave-vector to that of periodic absorption grating. Due to the steep atomic dispersion,
 this condition can be satisfied by small detuning of the signal light, in direct analogy
with other EIT-based processes \cite{harris-phystoday}.

In the case when the forward signal field propagates in the presence of only one (FD) control field,  
Eq.~(1) has a simple solution   $\Psi_+(z,t) = \Psi_+(z- c \tau(t), 0)$, which describes a coherent 
deceleration as shown in Figure 1C. Specifically, when the FD control beam is turned off, the polariton 
is stopped, the signal light vanishes, and a stationary spin excitation is created. Note that
the spatial length of the spin excitation $l$ is determined by the length of the compressed pulse
corresponding to  a product of the input pulse duration and the initial group velocity $v_g^0$. 
Finally, when the FD beam is turned back on, the propagating signal pulse is re-created.

When both control beams are turned on simultaneously, forward and backward signal components are both 
generated.   The electric field amplitudes for both forward and backward signal light are proportional to 
the amplitudes of the corresponding control fields, as required by the pulse matching 
condition \cite{harris-phystoday,match}.
Interesting insight into the propagation dynamics can be gained by considering the dispersion relation 
associated with Eqs.~(1,2):
 $\omega = - c k (\xi [\alpha_+-\alpha_-] - i k)/(\xi - i k [\alpha_+-\alpha_-])$, 
where $\omega$ is the Fourier frequency corresponding to the scaled time $\tau$ and $k$ is the spatial
wave-vector of the signal pulse envelopes. For small wave vectors and a large absorption coefficient 
this corresponds to a linear dispersive medium with a group 
velocity $v_g = c \times (|\Omega_+|^2 -|\Omega_-|^2)/g^2 N$. Thus, the time-domain solution describes 
the coupled motion of the $\Psi_{\pm}$ envelopes at a group velocity proportional to the intensity 
difference between the two control fields.  When the control Rabi-frequencies are identical, the group 
velocity $v_g$ vanishes and a stationary pulse of light is created as illustrated in Figure 1D. The 
electric field amplitudes of two signal components are then equal
${\mathcal E}_+ ={\mathcal E}_- $, corresponding to sinusoidally modulated signal intensity, 
and the dispersion relation becomes $\omega = i c k^2/\xi$. In the time-domain this corresponds to  
a slow spreading of the stationary pulse at a rate $\delta l/l \sim c \tau/(\xi l^2)$, which determines 
the maximal holding time. Hence, in an optically dense medium ($\xi l\gg 1$) a stationary photonic 
excitation can be controllably created.

Our experimental apparatus used to demonstrate this effect is shown in
Figure~2A. A magnetically shielded 4 cm long $^{87}\mbox{Rb}$ vapor cell is maintained at  
T $\sim 90$~${\rm ^{o}C}$ (atom number density is $10^{12}-10^{13}$~${\rm cm}^{-3}$).  Long-lived 
hyperfine sublevels  of the electronic ground state  $ S_{1/2}$ with $F=1,2$
are used as the storage states $|g \rangle, |s \rangle $, respectively,
coupled via the excited state $P_{1/2}$. The hyperfine coherence time is limited by atomic
diffusion out of the beam volume, and is enhanced with a Ne buffer gas of 6 torr
(resulting in spin-coherence lifetimes in  the $\mu$s to ms range \cite{kash99,phillips01}). Note that
the Doppler-shifts due to atomic motion do not directly affect the spin coherence when the control 
and the signal beams propagate colinearly in forward and backward pairs.

We first consider continuous wave (CW) excitation of the atomic Rb. Without the control beams the atomic
medium is completely opaque to the forward signal beam. When the FD control field is present, a sharp,
few 100 kHz-wide resonance appears in the transmission spectrum of the signal beam, corresponding to EIT 
(curve (i) in Fig.~2B). Turning on the BD beam in the CW regime greatly reduces the EIT transmission 
(curve (iii)), while at the same time generating a reflected signal beam that is detected in the backward 
direction (curve (ii)). The peak reflection intensity is a substantial fraction  (up to $\sim 80\%$) 
of the input signal beam.

These results demonstrate the possibility of coherent control of light via
simultaneous driving of the medium with FD and BD beams.  Specifically, the signal light cannot propagate 
in the modulated EIT medium,  but instead of
absorbing the signal the medium  reflects it as a high quality Bragg mirror. This effect is analogous to 
that predicted theoretically in \cite{axel}.
However, the broad lineshape of the curve (ii) indicates that periodic modulation of the absorptive rather 
than dispersive properties lays at the origin of the observed Bragg reflection \cite{kogelnik}. We note, 
in particular, good agreement between the present experimental results and a theoretical model based on the 
resonant EIT medium  with FD and BD control fields (Figure 2C). At the same time, qualitatively different 
lineshape is expected for dispersive Bragg gratings \cite{axel,slusher}.

Turning  to experiments with pulsed light,
we first map the input signal pulse onto an atomic coherence of the Rb atoms  \cite{liu01,phillips01}. 
This procedure corresponds to Fig. 1C and its experimental observation is shown
by curve (i) in Figure 3A. The atoms are first optically pumped into the lowest state. 
A  Gaussian-shaped signal pulse of about 5 $\mu$s duration then enters the medium 
where it is slowed to $v_g^0 \sim 6$km/s. The FD control beam is subsequently turned off. 
As a rule a fraction of the signal pulse leaves the cell before that, leading to the first peak 
of the curve (i). When the FD control field is turned back on, the stored atomic excitation is converted
back into light which is detected in the forward direction (second peak of the curve (i)).
The amplitude of the retrieved light decays with increasing storage interval with a characteristic time 
scale of about $20~ \mu$s, corresponding to decay of the hyperfine coherence caused by atomic diffusion. 
Similar experimental  results are obtained  by detecting the signal light in the backward direction
when the stored  coherence is retrieved with the BD beam \cite{sashaback}.

We next consider the retrieval of the atomic excitations by simultaneous
application of  the FD and BD control beams.
When the intensities of the beams are carefully adjusted,
the output signal pulses in both forward and backward directions are greatly suppressed
(curves (ii) and (iii) in Figure 3A). Both channels exhibit small leakage. We attribute the first peak 
to photons retrieved near the cell 
boundaries, which do not experience sufficient Bragg reflections to be trapped efficiently.  The 
long tail is likely due to a slow spreading of the stored pulse. When the BD beam is turned off 
the released pulse  is detected in the forward channel (curve (ii)).
The presence of signal light inside the cell during the simultaneous application of the two control beams  
was verified directly  by monitoring fluorescence from the side of the cell (Figure 3B). For times when the
signal output in forward and backward directions is greatly suppressed, we observed
significant enhancement of the signal light fluorescence  (curve (iii) in Figure 3B),  due to residual 
atomic absorption.

These observations provide evidence for controlled conversion of the stored atomic coherence into a 
stationary photonic excitation  in the cell. Note, in particular,  that the
magnitude of the  fluorescence drops sharply after the BD pulse is turned off. This drop is followed by a 
gradual decay associated with the exit of the slow pulse from the medium. This behavior is in a
qualitative agreement with our simple model, that predicts the light intensity in the stationary pulses 
to be double of that in the slowly propagating pulse.  
 As shown in Figure 3C, the magnitude of the released pulse decreases exponentially with 
increasing trapping time with a characteristic time constant of about
7 $\mu$s. Note that  only a part of this decay is due to the hyperfine coherence decay. Other decay 
mechanisms include spreading of the stationary pulse, as well as  imperfect  EIT. We anticipate
that improvements in efficiency can likely be achieved, e.g., by initial optical
pumping into a single atomic sublevel, using an atomic system with larger level
spacing or sharper absorption lines of cold atom clouds.

We finally outline  possible avenues opened by the present work. First, 
we note that our procedure is based on a passive medium and 
in the ideal limit is not accompanied  by optical loss or gain
and hence avoids the associated noise. We therefore
anticipate that our method preserves the quantum states  
of light pulses. (This is in contrast, e.g. to Bragg gratings
based on gain modulation \cite{kogelnik}.) Second,
although the present work demonstrates stationary light localization and storage in one dimension,  
it should be possible to controllably localize and guide  stationary photonic pulses in three spatial 
dimensions by using control beams with properly designed wavefronts. Third, controlled conversion of 
propagating light into stationary light pulses opens interesting possibilities for enhanced non-linear 
optical processes by combining the present technique with  the resonant enhancement of non-linear optics 
via EIT \cite{harris-nlo,atac-nlo,hemmer95}.
This combination may enable controlled interactions involving quantum few-photon
fields \cite{harris-hau,lukin-ima,gershon,tombessi} analogous to those feasible in cavity QED \cite{mabuchi}. 
Finally, extension of the present ideas to other systems might be possible using, for
example, dynamic modulation of  photonic bandgap materials.

{\bf Acknowledgments} We thank A.~Andre, M.~Eisaman, L.~Childress, C.~van der Wal, R.~Walsworth,
S.~Zibrov and T.~Zibrova for useful discussions, experimental help and comments on the manuscript.
This work is supported by the NSF,  the David and Lucille Packard Foundation and the Alfred Sloan 
Foundation. Partial support by the DARPA and the ONR (DURIP) is also acknowledged.






\newpage
   \begin{figure}
   \begin{center}\leavevmode
   \includegraphics[width=82mm]{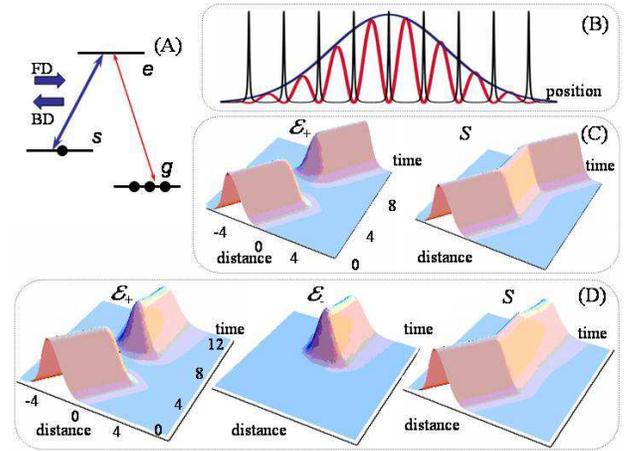}
    \caption{Physics of stationary pulses of light.
(A) Schematic of the atomic system.
The signal beam (red arrow) is resonant with the  transition from the ground to the excited state 
($|e\rangle$) and the control fields (blue arrows) are tuned to resonance between the $|s\rangle$ state 
and the excited state. (B) Schematic 
illustration of the spatial variation of the signal field absorption (black line), and the electric field 
(red line) in the stationary pulse in the medium along the direction of the control beam.  
Blue line represents initial atomic spin coherence. 
(C) Storage of the weak signal pulse in Raman coherence. Calculated evolution
is shown for the forward signal pulse and atomic spin wave amplitudes as a function of distance (in the units of compressed pulse length $l$) and time (in units of $l/v_g^0$). While the forward pulse
propagates  the FD beam is turned off ( at the time $t = 4 l/v_g^0$) and the pulse is stored in a stationary spin coherence. At the time $t =8 l/v_g^0$ the FD field is turned back on re-creating
the  propagating pulse. Calculations are based on Eqs.(1,2) with Gaussian pulses and
$\alpha_+ =1, \alpha_- =0$.   Figure
(D) shows calculated evolution of the forward and backward signal components and atomic spin wave amplitude for the case when the stored coherence is illuminated (at the time $t =8 l/v_g^0$) by FD and BD  beams with equal Rabi-frequencies. Here stationary signal pulses are created.
The main contributions to the dynamics include a slow spreading of the amplitudes as well as
small shifts (on the order of $1/\xi$) of forward and backward
components in the corresponding directions relative to the atomic coherence.
 Calculations are based on Eqs.(1,2) with
$\alpha_+ = \alpha_- =1/2,\xi l= 10$. All simulations assume that  $v_g\ll c$,
in which case the polaritons are mostly atomic and amplitude of the spin coherence
does not change significantly while the pulse is in the medium.  Amplitudes of
fields and coherence are normalized to their initial value at $t=0$.
}

   \label{fig:system87}
  \end{center}
   \end{figure}
   \begin{figure}
   \begin{center}\leavevmode
   \includegraphics[width=82mm]{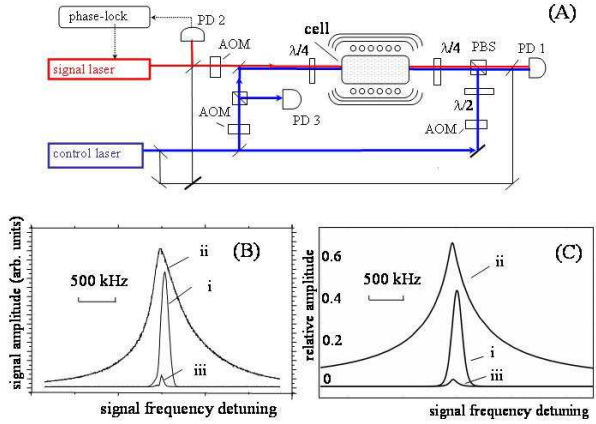}
    \caption{Experimental setup (A) and results of CW experiments (B). The signal field is generated by an extended cavity diode laser. The control beam from a locked Ti:Sapphire laser
is split to provide both FD and BD control
fields. The signal laser is phase-locked to the control laser with
a frequency offset corresponding to the hyperfine splitting of $^{87}$Rb
($6.84$ GHz). All three fields have  circular polarizations in the
cell but their intensities are controlled independently by acousto-optic modulators (AOMs).
The beams  overlap inside the cell at an angle of $\sim10^{-4}$ ~$\mathrm{rad}$.    PD1 and PD3 are fast photodiodes used for heterodyne detection of the the signal 
amplitude using a reference beam (black line). These  are
followed by a spectrum analyzer running in zero span mode (the detection
bandwidth is 3 MHz).  The CW power of
the signal beam is $250~\mu\mathrm{W}$, while the FD and BD powers are independently adjusted
between  $8.0~\mathrm{mW}$ and $40~\mathrm{mW}$. The beam size
inside the cell is about $2~\mathrm{mm}$.
     (B)  Curve (i) represents the EIT signal transmission when
the FD beam is on and  the BD beam is off. Transmission is approximately $50 \%$ at EIT resonance. Curve 
(ii)  is the reflected signal  and curve (iii) is the signal transmission when both FD and BD fields are on.   
(C) Corresponding theoretical simulations of the transmission and  reflection of signal field from the medium
 composed of atoms in Figure 1A. Parameters  correspond to atomic Rb, $\Omega_+ =\Omega_- = 15~$MHz,  
the spin decoherence rate of $3$ kHz, and  medium length corresponding to resonant attenuation of 
$e^{-15}$. The small frequency shift in the reflection resonance accounts for wave-vector matching to 
satisfy Bragg condition.
         }
   \label{fig2}
\end{center}
  \end{figure}
   \begin{figure}
   \begin{center}\leavevmode
   \includegraphics[width=95mm]{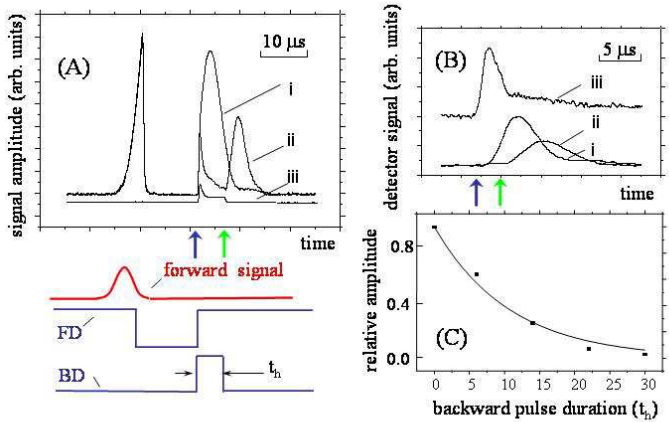}
    \caption{Results of pulsed experiments. (A) Detected output signals from the Rb cell.
Curve (i) is the signal detected in the forward direction resulting from pulse storage in a spin coherence. The left
peak represents the fraction of the signal pulse that leaves the cell before the trapping has begun.
(We were able to suppress this fraction to $\sim 10~\%$.)
It is delayed by $\sim 10~\mu\mathrm{s}$ compared to the original pulse (see timing diagram below).
The right peak in curve (i) appears only after the FD control 
is turned on, and thus
represents the stored and retrieved signal. Curve (ii) is the same except  the FD and BD  beams
are both turned on  as shown on the timing diagram and marked by the blue arrow.
Curve (iii) is the  signal in the backward direction  under the same conditions. (This
curve is plotted on a different scale with the peak signal  about a factor of five weaker
than that on curve (ii)). The pulse is released
in the forward direction when the BD control is turned off  (green arrow). On the timing diagram,
the rise  time edges of the control pulses
are about 0.1~$\mu\mathrm{s}$.  Note that the frequencies of the two control fields do not need to be exactly equal.  This is 
in agreement with theory and was verified  by obtaining results
similar to those in curve (ii)  with BD beam shifted by 80 MHz from FD.  (B) Rb fluorescence measured  at the side of the
cell. Curve (iii) is fluorescence associated with signal light during the release with both FD and BD 
beams on. Background fluorescence associated with control beams is subtracted. Curves (i) and (ii) (shown for reference) correspond to the same signals as curves (i) and (ii) in Fig.3A. Note that  fluorescence measurements are carried out under  conditions  differing from those for other data (including Fig.3A) since the
photodetector inserted inside the magnetic shields  introduces
stray magnetic fields resulting in shortened spin coherence and storage times; this measurement is also 
detection-bandwidth limited. (C) Dependence of the released signal pulse magnitude on the BD pulse duration $t_h$. 
    }
   \label{fig:rampulse}
   \end{center}
   \end{figure}

\end{document}